\documentclass[9pt,shortpaper,twoside,web]{ieeecolor}

\usepackage{generic}
\usepackage{cite}
\usepackage{amsmath,amssymb,amsfonts}
\usepackage{algorithm,algorithmic}
\usepackage{graphicx}
\usepackage{subfig}
\usepackage{MnSymbol,bm}
\usepackage{booktabs}
\usepackage[colorlinks,urlcolor=blue,linkcolor=blue,citecolor=blue]{hyperref}
\usepackage{multirow}
\usepackage{textcomp}
\hypersetup{hidelinks=true}
\usepackage{verbatim}
\usepackage{xcolor}
\usepackage{soul}
\usepackage{tikz} \usetikzlibrary{arrows.meta,positioning}

\newcommand{\orcid}[1]{\href{https://orcid.org/#1}{\includegraphics[width=8pt]{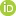}}}

\def\BibTeX{{\rm B\kern-.05em{\sc i\kern-.025em b}\kern-.08em
T\kern-.1667em\lower.7ex\hbox{E}\kern-.125emX}}

\begin{document}

\title{Is User Calibration Still Necessary for Continuous Myoelectric Control? A Closed-Loop Comparison of Calibration-Free Controllers}

\author{Amir~Reza~Hariri, Kevin~B.~Englehart\orcid{0000-0003-4525-1121}, and~Erik~Scheme\orcid{0000-0002-4421-1016}, 
\thanks{This work was supported in part by the Natural Sciences and Engineering Research Council of Canada (NSERC).}
\thanks{A. R. Hariri, K. B. Englehart, and E. Scheme are with the Institute of Biomedical Engineering and the Department of Electrical and Computer Engineering, University of New Brunswick, Fredericton, NB E3B 5A3, Canada.}
\thanks{Corresponding author: Erik Scheme (e-mail: escheme@unb.ca).}
}

\maketitle

\begin{abstract}

The need for user-specific calibration remains an often-cited barrier to widespread adoption of myoelectric control. Cross-user models trained on large multi-user datasets can now recognize gestures from unseen users without calibration, but they have been evaluated almost exclusively offline, leaving open whether calibration is still needed in closed-loop control. This study compared six calibration-free controllers, trained on 306 users of the EMG-EPN-612 dataset and used without any calibration data from the test participant, against three calibrated controllers in a closed-loop Fitts' law target-acquisition task with sixteen participants. The calibrated controllers achieved higher offline balanced accuracy (90.6\% against 73.1\% for the best controller of each paradigm), but completion rates did not differ significantly between paradigms, and the highest completion rate was achieved by a calibration-free controller. Calibrated controllers moved faster on successful acquisitions, partly because of their individually fitted speed mappings, though not significantly after correction for multiple comparisons. However, every calibration-free controller produced more direct and stable trajectories than every calibrated controller, even after adjusting for cursor speed. Offline accuracy did not predict completion rate across controllers or within participants, and across paradigms higher offline accuracy was associated with poorer trajectory quality in 15 of 16 participants. Without any calibration data from the end user, a model trained across a large population matched calibrated control in completion rate and exceeded it in trajectory quality. The offline accuracy gained through calibration should therefore not be taken as evidence of a closed-loop benefit.

\end{abstract}

\begin{IEEEkeywords}
Electromyography,
myoelectric control,
EMG,
cross-user,
calibration-free control,
human-machine interaction,
prosthetic control,
Fitts' law.
\end{IEEEkeywords}

\section{Introduction}
\label{sec:introduction}

Surface electromyography (EMG) provides a non-invasive estimate of motor intent and has long been used to control powered upper-limb prostheses~\cite{scheme2011electro,englehart2003robust}. More recently, the same signal has been applied to general human-computer interaction (HCI), including wearable input, cursor control, and augmented- and virtual-reality interfaces~\cite{saponas2008demonstrate,saponas2009enable,ctrl2024generic}. Most systems in both domains follow a standard pattern-recognition framework in which short windows of multichannel EMG are classified into motion classes that drive a device~\cite{englehart2003robust,scheme2011electro}. When these decisions are made continuously and combined with proportional control of speed, the user can start, steer, and stop movement in real time, but every misclassified window has a direct and visible effect~\cite{scheme2013fitts,hahne2014linear,jiang2014intuitive}. Errors therefore appear not only as incorrect gestures, but also as jitter, unintended motion, overshoot, and corrective movements. 

For most myoelectric systems, a persistent barrier to deployment is the need for user-specific calibration. Because muscle geometry, subcutaneous tissue, skin impedance, electrode placement, and contraction strategy all differ between individuals, a model trained on one user does not generalize well to another. Consequently, each new user must first provide labelled training data, typically through screen-guided training, before the system can be used~\cite{scheme2011electro,campbell2021deep}. Signal non-stationarity and confounding factors mean that this procedure must often be repeated across days, after donning and doffing, and with changes in limb position or perspiration~\cite{hwang2017real,campbell2021deep,vidovic2016improving}. In prosthetics, rejection of upper-limb devices remains common, at approximately 26\% for body-powered and 23\% for electric devices in a 25-year survey~\cite{biddiss2007upper} and 44\% in a more recent cohort~\cite{salminger2022current}, with control difficulty and training burden among the cited reasons. Reducing or eliminating calibration is therefore a key step toward both clinical adoption and consumer-scale HCI.

This need has motivated cross-user models (also described as user-independent, subject-independent, calibration-free, or zero-shot), in which a model trained on an existing population dataset is applied to a new user. Approaches differ in how much they rely on data from that user, though. Transfer learning and fine-tuning adapt a population model using a small amount of labelled data from the new user~\cite{ulysse2017transfer,coteallard2019deep,wu2023transfer}, and domain adaptation reduces the distribution shift between the training population and the end user~\cite{li2024across}. Zero-shot models, in contrast, are applied without any adaptation~\cite{campbell2021deep,eddy2024big}. Until recently, accuracy without adaptation was poor, falling below 50\% in some settings~\cite{campbell2021deep} and reaching 68\% for continuous cross-user recognition before transfer learning raised it to 92.3\%~\cite{xu2024crossuser}.

Larger datasets have begun to change this picture, though. Using the 612-user EMG-EPN-612 dataset~\cite{epn}, Eddy \emph{et al.} reached 93.0\% zero-shot gesture recognition accuracy on unseen users~\cite{eddy2024big}, and wrist-EMG-based temporal models have reached 78.2\% zero-shot, rising to 91.6\% with a single calibration repetition~\cite{botros2025zeroshot}. Industry-generated datasets now target unseen-user generalization at even larger scales~\cite{salter2024emg2pose,sivakumar2024emg2qwerty}. However, results of this kind typically classify complete gestures or accumulate evidence over windows well beyond the 200~ms decision interval used in continuous control, so they do not translate directly to the low-latency case. Many cross-user studies also address discrete gesture classification~\cite{eddy2024big,yang2024emgbench}, relying on temporal profiles to mitigate inter-user differences in EMG, leaving continuous cross-user control comparatively unexamined.

These results are also almost exclusively offline, obtained by classifying prerecorded data from held-out users~\cite{eddy2024big,coteallard2019deep,yang2024emgbench,botros2025zeroshot}. Research has repeatedly shown, however, that offline accuracy may be an unreliable predictor of closed-loop performance~\cite{Lock2005Realtime,ortizcatalan2015offline,Lv2019,Nawfel2021Multivariate}. Algorithms that differ significantly in offline accuracy have produced comparable online control, and some users have controlled a system effectively despite offline accuracies below 40\%~\cite{jiang2014accurate,hwang2017real}. The same dissociation has been observed in users with stroke and spinal cord injury~\cite{lu2019offline}, and a scheme that began with no user data and adapted online achieved higher Fitts' law throughput than a screen-guided baseline despite lower offline accuracy~\cite{campbell2024screen}. These findings motivated goal-oriented evaluations such as the Target Achievement Control Test~\cite{simon2011target} and Fitts' law target-acquisition tasks~\cite{scheme2013fitts,soukoreff2004towards}, which measure how well a user can complete a control task rather than how well a classifier labels data.

All of this evidence, however, was obtained with calibrated controllers. Closed-loop evaluation of a cross-user decoder remains rare. CTRL-labs reported human-in-the-loop performance for a generic decoder requiring no per-user calibration, including a median of 0.66 target acquisitions per second in continuous navigation and handwriting at 20.9 words per minute, with personalization improving handwriting decoding by a further 16\%~\cite{ctrl2024generic}. That work used proprietary hardware, though, and did not compare its generic decoder with calibrated alternatives in closed loop, so to the best of our knowledge, no matched comparison of cross-user and calibrated controllers has been reported. It is therefore not known whether the reported offline advantage of calibrated models carries over to closed-loop control. This matters in practice, because selecting models on offline accuracy alone may discard cross-user controllers that would perform well in closed loop, or may overstate the benefit of calibration.

This study therefore compared six calibration-free cross-user controllers against three calibrated controllers under a common closed-loop protocol. The cross-user controllers were trained on a user-disjoint partition of the EMG-EPN-612 dataset~\cite{epn} and received no data from the online participants, and all but one of the nine controllers shared a common convolutional network backbone. Performance was evaluated using a two-dimensional ISO~9241-9 Fitts' law target-acquisition task with proportional velocity control~\cite{fitts1954information,soukoreff2004towards,scheme2013fitts}. For every controller, offline balanced accuracy was computed alongside online completion rate, throughput, path efficiency, overshoot, direction changes, and stopping distance. These measures allowed us to examine whether user-specific calibration still provides a closed-loop benefit when a cross-user model is trained on hundreds of users, and whether offline accuracy, the measure by which cross-user and calibrated models are usually compared, predicts that closed-loop benefit.

This work correspondingly makes three main contributions:
\begin{enumerate}
\item The first matched closed-loop test, to our knowledge, of whether user-specific calibration is still needed, comparing six calibration-free and three calibrated controllers within the same participants and task.
\item Without any data from the user, calibration-free control matched calibrated control in completion rate and produced more direct and stable trajectories than every calibrated controller, so calibration provided no significant closed-loop benefit.
\item Calibration raised offline accuracy by 17 to 31 percentage points, yet this gain did not carry over to completion rate and came with poorer trajectories in 15 of 16 participants, showing that offline accuracy, the usual basis for comparing calibrated and cross-user models, can point in the wrong direction.
\end{enumerate}

\section{Methods}
\label{sec:methods}
\subsection{Participants}
\label{sec:participants}

Sixteen able-bodied participants (aged 21--45 years, mean $26.3 \pm 6.1$; 10 male, 6 female; 15 right-handed, 1 left-handed) provided informed consent and completed the study in a single session as approved by the University of New Brunswick Research Ethics Board (REB 2024-082). They rated their prior experience with myoelectric control as none (6), low (4), medium (3), or high (3).  Participants wore an 8-channel Myo armband (Thalmic Labs), sampled at 200~Hz, on the right forearm in the same orientation used to collect the EMG-EPN-612 dataset~\cite{epn}.

\subsection{Control Paradigms}
\label{sec:paradigms}

Nine controllers were compared in this study. Six were calibration-free, meaning that they were trained only with users from the EMG-EPN-612 dataset and used without any calibration data or adaptation from the participant. The other three were calibrated using data collected from each participant at the start of the session.

The cross-user models were trained on EMG-EPN-612, which contains recordings from 612 users performing six gestures with the Myo armband~\cite{epn}. The pinch gesture was excluded, leaving five classes (hand open, hand close, wrist flexion, wrist extension, and no motion) that aligned with the cardinal directions of the target acquisition task described in Section \ref{sec:protocol}. A fixed user-disjoint partition was used throughout development, with 306 training users (users 1--306), 26 validation users (307--332), and 280 test users (333--612). All hyperparameter and model selection used the validation users only, and no data from the test users or the online participants were used during development. EMG data were segmented into 200~ms windows with a 25~ms increment.

All calibration-free controllers, and the two fine-tuned calibrated controllers, used a Multi-Horizon Convolutional Neural Network (MHCNN) designed for short-window EMG classification. The network takes an $8 \times 40$ window (eight channels and 200~ms at 200~Hz) and passes it through three parallel convolutional branches with a kernel size of eight and dilation factors of 1, 2, and 4, giving effective receptive fields of 40, 75, and 145~ms. The branch outputs are concatenated and fused by a further convolutional layer, and adaptive average pooling and fully connected layers then produce a 128-dimensional embedding that a linear layer maps to the five classes. The network has 89,189 trainable parameters.

The MHCNN was selected based on a zero-shot benchmarking of seven architectures, including linear, recurrent, and convolutional models operating on either hand-crafted features or raw EMG, each trained on the 306 training users and tested on the 280 test users. It achieved the highest balanced accuracy (79.3\%, against 77.8\% for the next-best model and 67.3\% for LDA), and its advantage over every other architecture was significant (Friedman test with Holm--Bonferroni-corrected pairwise comparisons). Because the conventional raw-input and feature-based models of similar capacity performed comparably, this advantage appears to come from the multi-scale receptive field rather than from the input representation. A single backbone was then used for all cross-user controllers, so that differences between them could be attributed to the training or preprocessing intervention rather than to the architecture.

\subsubsection{Calibration-Free Controllers}

The Base model was the MHCNN trained with standard cross-entropy loss. Each of the other five controllers differed from Base by a single intervention applied at a different point in the pipeline, so that any difference in performance could be attributed to that intervention alone.

\begin{itemize}
    \item \textbf{Segmented} modified the training data. The active interval of each gesture repetition was detected from the smoothed energy envelope of its highest-energy channel, using a threshold of 25\% of the envelope's range, and windows outside this interval were discarded. Rest repetitions were left unchanged. This removed gesture onsets and offsets, where residual or rebound activity often makes the recorded signal inconsistent with its assigned label. Segmentation was only applied to the training data, to maintain a fair test comparison with other models.

    \item \textbf{RunNorm} modified the input at inference. The model was trained on EMG standardized with each training user's own channel statistics. At inference, where these statistics are not available, RunNorm maintained a causal, cumulative estimate of the mean and variance of each channel, initialized from the training-population statistics and updated with every window observed. Normalization therefore began from the population priors and moved toward the current user's statistics as use continued, without any dedicated calibration data. The estimates were reset at the start of each session.

    \item \textbf{Proto} and \textbf{Triplet} modified the learned embedding. Proto added a prototypical attraction-repulsion term~\cite{snell2017prototypical} and Triplet added a cross-user triplet loss~\cite{hermans2017defense}, each as an auxiliary loss alongside cross-entropy. Both aimed to produce more compact and separable gesture classes without using user-specific information.

    \item \textbf{RestLoss} modified the error weighting. The cross-entropy loss of each window was doubled whenever the model predicted an incorrect active class, and doubled again if the true class was rest. False activations during rest were therefore weighted four times as heavily as missed activations (active windows predicted as rest), because during closed-loop control, an incorrect movement is more costly to the user than no movement at all~\cite{8516279}.
\end{itemize}

Testing combinations of these interventions online was not feasible within a single session, so each was evaluated on its own.

\subsubsection{Calibrated Controllers}

The three calibrated controllers were trained using up to five repetitions per gesture collected from the participant at the start of the session (Section~\ref{sec:protocol}).

\begin{itemize}
    \item \textbf{Within-5} was trained from scratch on five repetitions per gesture. It used a convolutional network on hand-crafted wavelet-energy features (CNN-HCF) rather than the MHCNN, because feature-based models are more data efficient when trained on limited data from a single user. In within-user evaluation on the 280 EMG-EPN-612 test users for the same five-repetition problem, CNN-HCF reached 86.5\% balanced accuracy, ahead of an MLP (85.4\%), an LSTM on the same features (81.7\%), and the MHCNN (78.9\%). Within-5 therefore represents the strongest from-scratch calibrated model available at this calibration budget.

    \item \textbf{FT-1} and \textbf{FT-5} started from the weights of Base and were fine-tuned on one and five repetitions per gesture, respectively, with all layers updated~\cite{ulysse2017transfer,li2025deep}. Starting from Base meant that these controllers differed from Base only by the addition of calibration data. In the same within-user evaluation on the 280 EMG-EPN-612 test users, fine-tuning raised MHCNN balanced accuracy to 86.4\% with one repetition and 91.5\% with five, exceeding every model trained from scratch, so FT-5 represents the strongest calibrated reference available.
\end{itemize}

All three were trained with the Adam optimizer at an initial learning rate of $5\times10^{-4}$ and a batch size of 64, with the learning rate reduced by a factor of 0.6 after seven epochs without improvement in validation loss. Training ran for up to 300 epochs with early stopping after fifteen epochs without improvement, using the participant's sixth repetition for validation, and the checkpoint with the lowest validation loss was used online.

All nine controllers used the same armband, gesture set, and velocity-control scheme. They differed in how the proportional-control thresholds were set, however, because the calibrated controllers could use the participant's own data and the calibration-free controllers could not (Section~\ref{sec:cursor_control}).

\subsection{Experimental Protocol}
\label{sec:protocol}

\begin{figure}[h]
\centering
\resizebox{0.85\columnwidth}{!}{%
\begin{tikzpicture}[
    box/.style={draw, rounded corners=2mm, align=center, inner sep=4pt, font=\small, fill=gray!10},
    arrow/.style={-{Latex[length=2mm]}, thick}
]
\node[box, minimum width=3.0cm, minimum height=1.25cm] (epn) at (-2.2,0)
{\textbf{EMG-EPN-612}\\306 training users};
\node[box, minimum width=3.0cm, minimum height=1.25cm] (sgt) at (2.2,0)
{\textbf{Participants} ($n=16$)\\Screen-guided training\\6 repetitions per gesture};

\node[box, fill=blue!12, minimum width=3.0cm, minimum height=2.7cm] (cf) at (-2.2,-2.6)
{\textbf{Calibration-Free}\\[1mm]Base\\Proto\\RestLoss\\Triplet\\RunNorm\\Segmented};
\node[box, fill=green!12, minimum width=3.0cm, minimum height=2.7cm] (cal) at (2.2,-2.6)
{\textbf{Calibrated}\\[1mm]Within-5\\FT-1\\FT-5\\ \\ \\ };

\draw[arrow, blue!70!black] (epn.south) -- (cf.north);
\draw[arrow, green!50!black] (sgt.south) -- node[right, font=\footnotesize, align=left] {Reps 1--5} (cal.north);
\draw[arrow, dashed, gray!70!black] (epn.south east) -- node[pos=0.5, below, sloped, font=\scriptsize] {Base weights} (cal.north west);

\node[circle, fill=black, inner sep=1.2pt] (merge) at (0,-4.5) {};
\node[box, minimum width=8.1cm] (eval) at (0,-5.4)
{\textbf{Common Closed-Loop Evaluation} (randomized order)\\Fitts' law task, 3 difficulty levels, 36 acquisitions per controller};
\node[box, minimum width=8.1cm] (out) at (0,-6.9)
{\textbf{Outcomes}\\Offline accuracy (using Rep 6) $\bullet$ Reliability $\bullet$ Speed\\Throughput $\bullet$ Trajectory quality};

\draw (cf.south) |- (merge); \draw (cal.south) |- (merge);
\draw[arrow] (merge) -- (eval.north);
\draw[arrow] (eval.south) -- (out.north);
\end{tikzpicture}
}
\caption{Overview of the study design. Calibration-free controllers were trained only on EMG-EPN-612, whereas calibrated controllers also used five screen-guided repetitions from each participant. All nine controllers were then evaluated by each participant, in randomized order, in a common closed-loop Fitts' law task.}
\label{fig:study_overview}
\end{figure}
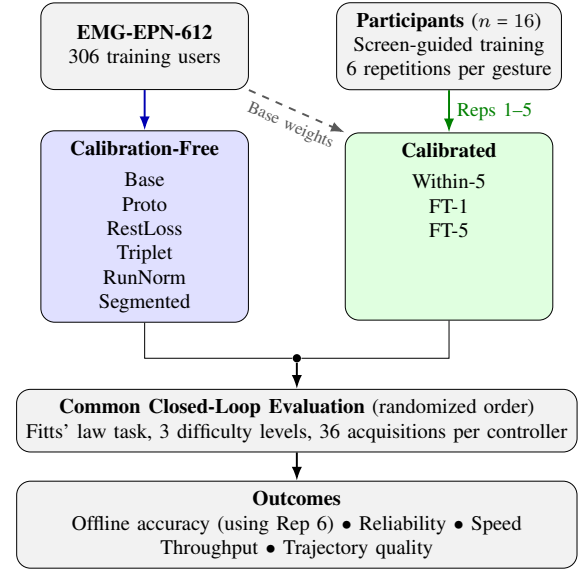

The study design is summarized in Fig.~\ref{fig:study_overview}. Participants first had up to five minutes of familiarization with the Base controller to learn the gesture-to-direction mapping and the proportional-control scheme. Base was used for familiarization because it is the same for every participant. Familiarizing with a calibrated controller could have allowed participants to develop strategies suited to their own model first, which would then have had to change when they switched to the calibration-free controllers. They then completed a screen-guided training session consisting of six repetitions of each gesture, each a three-second prompted contraction. The first five repetitions were used to train the calibrated controllers (FT-1 used a single repetition), and the sixth was held out to compute offline accuracy for all nine controllers. The sixth repetition was also used for early stopping of the calibrated controllers (Section~\ref{sec:paradigms}), so their offline accuracies are likely to be optimistically biased.  A budget of five repetitions was chosen to reflect the upper limit of calibration data typical of practical myoelectric systems.

All nine controllers were then evaluated online, in an order randomized independently for each participant to reduce learning and fatigue effects. Participants were blinded to the controller they were using, and rest breaks were given between controllers. Each participant completed the full Fitts' law protocol with every controller, giving 36 target acquisitions per controller, and the complete session took approximately one hour.

\begin{figure}[t]
\centering
\includegraphics[width=0.6\linewidth]{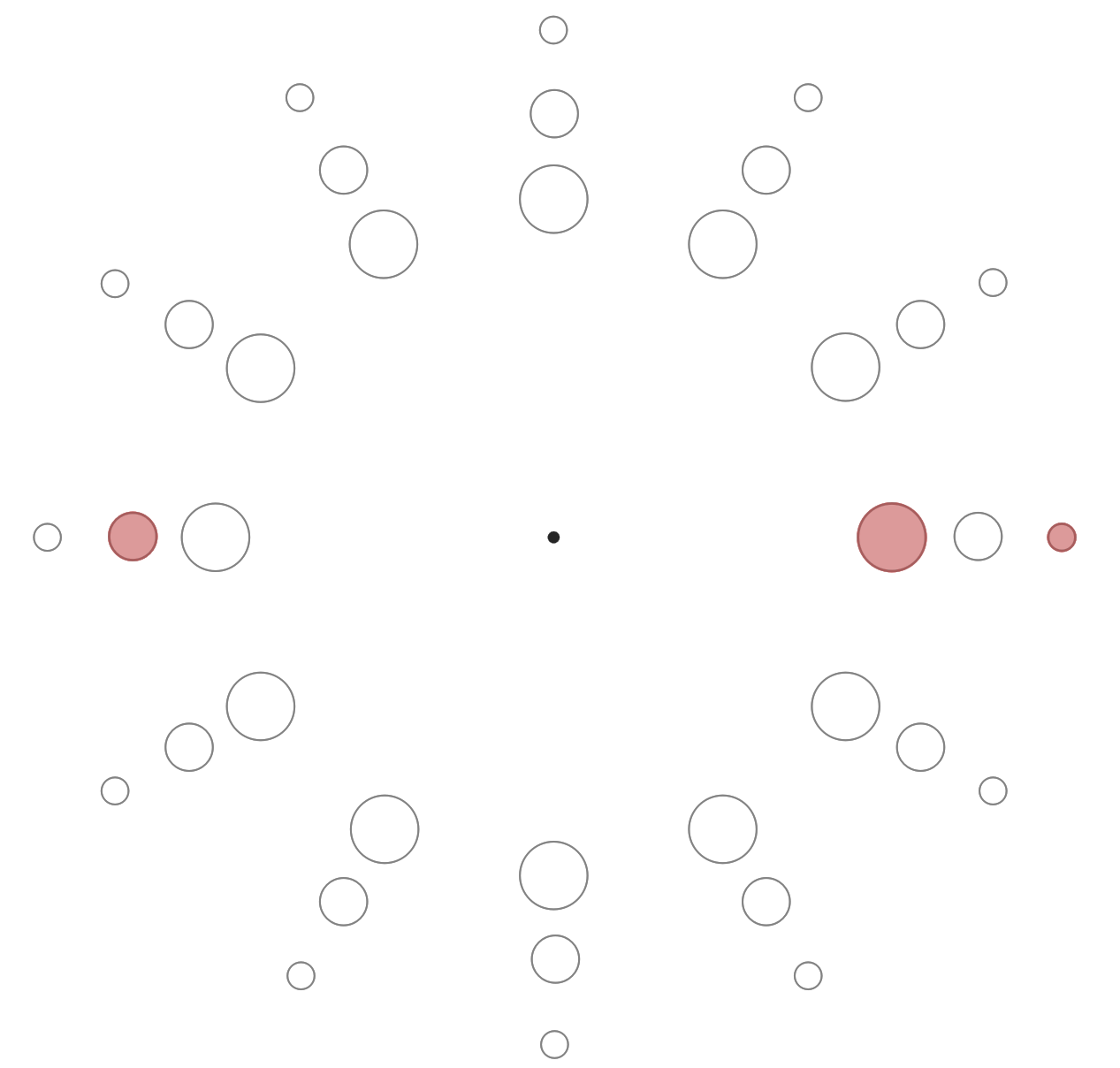}
\caption{Fitts' law task layout, with the three difficulty conditions overlaid. Each condition is a ring of twelve targets around a central start point, with larger targets at shorter distances in the easy condition and smaller targets at longer distances in the hard condition. Shaded circles show example targets. A target was acquired by holding the cursor inside it for 0.75~s.}
\label{fig:fitts_environment}
\end{figure}

Online performance was evaluated using a two-dimensional Fitts' law target-acquisition task with targets arranged in the multidirectional ISO~9241-9 ring~\cite{mackenzie1992fitts,soukoreff2004towards}, which has become a common framework for evaluating closed-loop myoelectric control~\cite{scheme2013fitts}. Participants used EMG to move a cursor to a sequence of targets (Fig.~\ref{fig:fitts_environment}), and task difficulty was quantified using the Shannon formulation of Fitts' law,
\begin{equation}
ID=\log_2\left(\frac{A}{W}+1\right),
\end{equation}
where $A$ is the movement amplitude and $W$ is the target width.

Three difficulty conditions were used, with indices of difficulty from 3.46 to 5.27~bits (Table~\ref{tab:fitts_conditions}). Each condition consisted of a ring of twelve targets presented in the standard multidirectional sequence. A target was acquired when the cursor remained inside it for a dwell time of 0.75~s, and trials not completed within 8~s were recorded as failures. These parameters were chosen during pilot testing to give completion rates of approximately 80\%, 65\%, and 50\% for the easy, medium, and hard conditions, respectively.

\begin{table}[hb]
\centering
\caption{Fitts' law target-acquisition difficulty conditions. $A$ denotes movement amplitude and $W$ denotes target width.}
\label{tab:fitts_conditions}
\small
\begin{tabular}{lccc}
\toprule
\textbf{Condition} & \textbf{$W$ (px)} & \textbf{$A$ (px)} & \textbf{ID (bits)} \\
\midrule
Easy   & 60 & 600 & 3.46 \\
Medium & 42 & 750 & 4.24 \\
Hard   & 24 & 900 & 5.27 \\
\bottomrule
\end{tabular}

\end{table}

\subsection{Cursor Control}
\label{sec:cursor_control}

Cursor movement used a conventional classification-based proportional control scheme~\cite{scheme2013fitts,scheme2014mnpc}, in which the predicted class set the direction of movement and EMG amplitude set its speed. The four active gestures were mapped to the four cardinal directions, and the no-motion class stopped the cursor. Movement intensity was computed as the channel-summed mean absolute value of each window,
\begin{equation}
m(\mathbf{x})=\sum_{c=1}^{8}\frac{1}{T}\sum_{t=1}^{T}|x_c[t]|,
\end{equation}
where $x_c[t]$ is sample $t$ of channel $c$ and $T$ is the number of samples in the window.

Intensity was mapped to speed through a clipped linear ramp with class-specific thresholds, following the motion-normalized approach of~\cite{scheme2014mnpc}, so that each gesture could reach full speed at an intensity appropriate to that gesture. For class $k$, let $r^k_{\min}$ and $r^k_{\max}$ be the minimum and maximum window intensity in the reference data for that class. The thresholds were
\begin{equation}
\theta^k_{lo}=0.9\,r^k_{\min}+0.1\,r^k_{\max},
\qquad
\theta^k_{hi}=0.3\,r^k_{\min}+0.7\,r^k_{\max},
\end{equation}
and the speed was
\begin{equation}
s=\min\!\left(1,\ \max\!\left(0,\ \frac{m(\mathbf{x})-\theta^k_{lo}}{\theta^k_{hi}-\theta^k_{lo}}\right)\right),
\end{equation}
so that intensities at or below $\theta^k_{lo}$ produced no motion and intensities at or above $\theta^k_{hi}$ produced full speed. Cursor velocity was $\mathbf{v}=s\,g\,\mu\,\mathbf{u}_k$, where $\mathbf{u}_k$ is the unit vector in the direction of class $k$, $g=20$~px per frame (at 60 frames per second) is the base gain, and $\mu$ is a paradigm-level velocity multiplier. For the calibrated controllers, $r^k_{\min}$ and $r^k_{\max}$ were taken from the participant's own calibration repetitions, which tuned proportional control to that participant's contraction levels. The calibration-free controllers could not use participant data and instead took these values from the EMG-EPN-612 training population.

Motion-normalized proportional control assumes that each class's speed mapping reflects the individual user's own contraction range~\cite{scheme2014mnpc}. A mapping derived from the pooled training data violates this assumption and is more conservative, because $r^k_{\max}$ is set by the strongest contractions in the whole population. This raises $\theta^k_{hi}$ above the value an individually calibrated user would have and compresses the usable range, which would slow calibration-free cursor movement regardless of classifier quality. To offset this, the calibration-free controllers were given $\mu=1.2$, against $\mu=1.0$ for the calibrated controllers. This value was not obtained through a formal search. It was set during informal pilot testing as the adjustment that brought cross-user cursor speed into a range comparable to the calibrated controllers, and it was fixed before data collection began. The size of the remaining mismatch was quantified after the study and is reported in Section~\ref{sec:paradigm_comparison}.

\subsection{Performance Metrics}
\label{sec:metrics}

\subsubsection{Offline Performance}

Offline performance was measured as balanced accuracy, the mean of the per-class recalls, on each participant's held-out sixth screen-guided repetition. Because all repetitions each contained one three-second contraction of each class, balanced accuracy is equal to overall accuracy here. 

\subsubsection{Online Performance}

The primary online outcome was completion rate (CR), the percentage of targets acquired before the 8~s timeout. Movement time (MT) was the time from target appearance to acquisition, excluding the 0.75~s dwell. Throughput (TP), which combines speed and accuracy into a single rate of information transfer, was computed as the mean of $\mathrm{ID}_i/\mathrm{MT}_i$ over trials~\cite{mackenzie1992fitts,soukoreff2004towards,scheme2013fitts}. Three versions of throughput are reported. Nominal throughput uses successful acquisitions only, and effective throughput replaces the nominal index of difficulty with an effective index computed from the observed spread of endpoints~\cite{soukoreff2004towards}. Penalized throughput, and likewise penalized movement time (MT$_P$), assigns the full timeout to each failed acquisition before averaging, so that failures are counted rather than ignored.

Trajectory quality was described by four measures. Path efficiency and the direction-change ratio were computed over all trials, and overshoot count and stopping distance over successful acquisitions. Path efficiency is the ratio of the straight-line distance from start to endpoint to the length of the path actually travelled, and reflects how directly the cursor moved. The direction-change ratio is the number of cardinal movement segments used to reach the target divided by the minimum number required, so that an ideal trajectory scores 1.0 and higher values indicate more corrections. Overshoot count is the number of times the cursor left the target after first entering it, and stopping distance is the maximum distance the cursor travelled past the target boundary after first entry. Higher path efficiency indicates better control, whereas lower values are better for the other three measures.

For the paradigm-level comparisons, the four measures were also combined into a single trajectory-quality composite. Each measure was standardized over all participant-by-controller cells, its sign was aligned so that higher values indicate better control, and the four standardized scores were averaged with equal weight,
\begin{equation}
Q_{s,m}=\frac{1}{4}\sum_{j=1}^{4}\sigma_j\,\frac{q_{j,s,m}-\bar{q}_j}{\mathrm{SD}(q_j)},
\end{equation}
where $q_{j,s,m}$ is measure $j$ for participant $s$ with controller $m$, the mean and standard deviation (SD) are taken over all participant-by-controller cells, and $\sigma_j=+1$ for path efficiency and $-1$ for the other three measures. The composite is expressed in standard-deviation units and is therefore meaningful only for comparisons within this dataset.

\subsection{Statistical Analysis}
\label{sec:stats}

All comparisons between controllers were paired, with each participant serving as their own control. Because of this pairing and the modest sample size ($n=16$), the Wilcoxon signed-rank test was used for every pairwise comparison. Effect sizes are reported as the matched-pairs rank-biserial correlation ($r_{rb}$), which accompanies the Wilcoxon test, together with Cohen's $d_z$ for comparison with studies that use parametric statistics. Contrasts were organized into three families:

\begin{itemize}
    \item \textbf{Family A} tested whether each calibration-free intervention (Proto, RestLoss, Triplet, RunNorm, and Segmented) improved on Base in completion rate, the trajectory-quality composite, and penalized movement time. Because the five comparisons share a common reference, $p$-values were Holm--Bonferroni corrected within each outcome.

    \item \textbf{Family B} tested whether the two paradigms differed overall. Each participant contributed one score per paradigm, computed as the unweighted mean over the controllers in that paradigm, and six outcomes were compared (completion rate, the trajectory-quality composite, movement time, penalized movement time, effective throughput, and penalized nominal throughput). Both uncorrected and Holm--Bonferroni-corrected $p$-values are reported, so that the effect of correction on each contrast is visible.

    \item \textbf{Family C} compared the strongest controller of each paradigm on the same six outcomes. The strongest controller of each paradigm was defined as the one with the highest mean offline balanced accuracy, which was measured on the held-out repetition before and independently of the online outcomes, so the selection was not influenced by the results being compared. The calibration-free controller with the second-highest offline balanced accuracy was also compared against the same calibrated controller to test whether the result depended on the choice of calibration-free controller. $p$-values were Holm--Bonferroni corrected across the six outcomes within each contrast.
    
\end{itemize}

To test whether the paradigm differences in trajectory quality could be explained by cursor speed, each trajectory measure was also analysed at the trial level using a mixed model (Poisson for overshoot count; linear for others) with paradigm, task difficulty, and mean trial speed as fixed effects and participant and controller as random intercepts (Section~\ref{sec:paradigm_comparison}). The correspondence between offline balanced accuracy and online outcomes was also measured across three levels. Spearman correlations were computed across controller means, across controllers within each participant (with the resulting coefficients tested against zero using the Wilcoxon signed-rank test), and across participants. A linear mixed model with participant as a random intercept was then used to separate the within-participant and between-participant effects of offline accuracy, with and without paradigm as a fixed effect.

\section{Results}
\label{sec:results}

The offline accuracy of each participant's held-out repetition is summarized in Table~\ref{tab:offline_results}. The calibrated controllers were more accurate than every calibration-free controller, and FT-5 reached the highest balanced accuracy (90.6\%), compared with 73.1\% for the strongest calibration-free controller, Triplet. FT-1 reached 84.4\% from a single calibration repetition. Five of the six calibration-free controllers fell within 2 percentage points of one another (71.1\% to 73.1\%), with Segmented second at 72.7\%.

RunNorm was the exception, at 59.9\%. It was the only controller whose output depended on the data it had already seen, so it interacted directly with the structure of the offline test. The held-out recording presented one three-second repetition of each class in a fixed order that began with no motion. The running statistics were therefore estimated almost entirely from rest before any active gesture was scored, and rest windows were then rescaled by statistics taken from rest itself, which biased the normalization toward the no-motion class. A single repetition per class was too short for the running statistics to stabilize. During offline development, by contrast, each EMG-EPN-612 test user provided 25 repetitions of each gesture, which was enough for the statistics to stabilize. Presenting the held-out windows in random order also raised balanced accuracy, but with so few windows available the size of the gain depended on the random seed and the resulting order. RunNorm's offline score therefore understates its performance because of the length of the test, which is consistent with its stronger standing online (Section~\ref{sec:reliability}).

\begin{table}[t]
\centering
\caption{Offline performance of the models evaluated in the online study. Cross-user models were trained on EMG-EPN-612 and evaluated zero-shot using each participant's held-out screen-guided training repetition. Calibrated models were trained or fine-tuned using participant-specific calibration data. Values are mean $\pm$ standard deviation across the sixteen participants. All controllers, including Segmented, were scored on the same unsegmented repetition.}

\label{tab:offline_results}
\small
\setlength{\tabcolsep}{10pt}

\begin{tabular}{lc}
\toprule
\textbf{Model} & \textbf{Balanced Accuracy (\%)} \\
\midrule
Base      & $71.3 \pm 12.1$ \\
Proto     & $71.1 \pm 12.1$ \\
RestLoss  & $71.2 \pm 11.6$ \\
Triplet   & $73.1 \pm 12.0$ \\
RunNorm   & $59.9 \pm 4.4$  \\
Segmented & $72.7 \pm 11.7$ \\
\midrule
Within-5  & $89.9 \pm 6.6$ \\
FT-1      & $84.4 \pm 9.5$ \\
FT-5      & $90.6 \pm 7.3$ \\
\bottomrule
\end{tabular}

\end{table}

\subsection{Online Reliability}
\label{sec:reliability}

The task produced the expected Fitts' law relationship, with movement time increasing approximately linearly with index of difficulty for every controller (grand $R^2$ from 0.91 to 0.99, and mean per-participant $R^2$ from 0.77 to 0.91).

Completion rate, the primary online outcome, is reported in Table~\ref{tab:online_outcomes}. Every calibration-free intervention improved on Base. Segmented reached the highest completion rate overall (72.7\%), followed by Triplet (71.0\%) and RestLoss (70.8\%), while RunNorm (69.4\%) and Proto (68.1\%) improved on Base by a smaller margin. This ordering differs from the offline ordering in Table~\ref{tab:offline_results}, most notably for RunNorm, which was the weakest calibration-free controller offline but closed much of the gap online. Its online advantage did not build up over the session, though. Relative to the mean of the other five calibration-free controllers at the same trial index, which holds condition order and practice constant, its completion-rate difference was $+0.8$ percentage points in the first half of a block and $-1.0$ in the second ($p=0.53$), with no reliable trend in penalized movement time ($p=0.90$) or path efficiency ($p=0.14$).

Despite their higher offline accuracy, the calibrated controllers did not achieve higher online reliability. Within-5, FT-1, and FT-5 completed between 67.2\% and 68.2\% of targets, below four of the six calibration-free controllers, and three calibration-free controllers exceeded 70\% without any participant-specific training.

\begin{table}[t]
\centering
\caption{Online outcome metrics, reported as mean $\pm$ standard deviation across the sixteen participants. Completion rate (CR) measures reliability, while movement time (MT) and throughput (TP) summarize speed and information transfer. Throughput is nominal throughput, and penalized metrics (subscript $P$) include failed acquisitions.}
\label{tab:online_outcomes}
\footnotesize
\setlength{\tabcolsep}{3pt}
\resizebox{\columnwidth}{!}{%
\begin{tabular}{lccccc}
\toprule
\textbf{Model} & \textbf{CR (\%)} & \textbf{MT (s)} & \textbf{MT$_P$ (s)} & \textbf{TP (b/s)} & \textbf{TP$_P$ (b/s)} \\
\midrule
Base      & $65.1 \pm 26.8$ & $4.16 \pm 0.87$ & $5.25 \pm 1.45$ & $1.13 \pm 0.35$ & $1.01 \pm 0.39$ \\
Proto     & $68.1 \pm 26.2$ & $4.17 \pm 0.82$ & $5.16 \pm 1.39$ & $1.11 \pm 0.35$ & $1.01 \pm 0.39$ \\
RestLoss  & $70.8 \pm 23.8$ & $4.04 \pm 0.85$ & $4.96 \pm 1.39$ & $1.18 \pm 0.36$ & $1.07 \pm 0.38$ \\
Triplet   & $71.0 \pm 26.2$ & $4.00 \pm 0.74$ & $4.96 \pm 1.41$ & $1.17 \pm 0.34$ & $1.06 \pm 0.40$ \\
RunNorm   & $69.4 \pm 23.1$ & $4.06 \pm 0.88$ & $5.05 \pm 1.35$ & $1.18 \pm 0.36$ & $1.06 \pm 0.40$ \\
Segmented & $72.7 \pm 25.5$ & $4.32 \pm 0.90$ & $5.06 \pm 1.39$ & $1.12 \pm 0.35$ & $1.04 \pm 0.38$ \\
\midrule
Within-5  & $67.9 \pm 24.4$ & $4.01 \pm 0.57$ & $5.09 \pm 1.20$ & $1.19 \pm 0.27$ & $1.04 \pm 0.32$ \\
FT-1      & $68.2 \pm 30.6$ & $4.03 \pm 0.90$ & $4.98 \pm 1.59$ & $1.23 \pm 0.37$ & $1.11 \pm 0.44$ \\
FT-5      & $67.2 \pm 27.4$ & $3.89 \pm 0.89$ & $4.97 \pm 1.53$ & $1.28 \pm 0.40$ & $1.12 \pm 0.45$ \\
\bottomrule
\end{tabular}%
}
\end{table}

\subsection{Reliability Across Task Difficulty} \label{sec:reliability_by_difficulty}

Completion rate declined with task difficulty for every controller (Table~\ref{tab:completion_by_difficulty}), as expected from the Fitts' law speed-accuracy trade-off. In the easy condition most controllers exceeded 80\% and the differences between them were small, but the controllers separated more clearly as difficulty increased. In the hard condition, Segmented and Triplet reached 56.2\% and FT-1 reached 54.2\%, whereas Base reached only 45.3\%. The advantage of the strongest calibration-free intervention, Segmented, over Base was larger in the hard condition (10.9 percentage points) than in the easy condition (4.1).

At the paradigm level, the calibration-free group led by 5.1 percentage points in the easy condition, and the two paradigms were within 0.6 points of each other in the medium and hard conditions. At every difficulty level, the best calibration-free controller completed more targets than the best calibrated controller.

\begin{table}[t]
\centering
\caption{Completion rate (\%) by Fitts' law difficulty condition. Values are averaged across the sixteen participants.}
\label{tab:completion_by_difficulty}
\small
\begin{tabular}{lccc}
\toprule
\textbf{Model} & \textbf{Easy} & \textbf{Medium} & \textbf{Hard} \\
\midrule
Base      & 81.8 & 68.2 & 45.3 \\
Proto     & 86.5 & 67.7 & 50.0 \\
RestLoss  & 88.5 & 75.0 & 49.0 \\
Triplet   & 88.0 & 68.8 & 56.2 \\
RunNorm   & 83.3 & 75.0 & 50.0 \\
Segmented & 85.9 & 76.0 & 56.2 \\
\midrule
Within-5  & 82.3 & 72.9 & 48.4 \\
FT-1      & 79.2 & 71.4 & 54.2 \\
FT-5      & 80.2 & 72.4 & 49.0 \\
\bottomrule
\end{tabular}

\end{table}

\subsection{Movement Time and Throughput}
\label{sec:speed}

Movement time and throughput describe how quickly targets were acquired, and their interpretation depends on whether failed acquisitions are counted. On successful acquisitions only, the calibrated controllers had the three highest nominal throughputs, and FT-5 had both the lowest movement time and the highest throughput (3.89~s and 1.28~b/s, Table~\ref{tab:online_outcomes}). Differences in movement time were smaller than those in throughput, though, and Triplet (4.00~s) was within 0.01~s of Within-5 and faster than FT-1.

The calibrated advantage shrank once failed acquisitions were charged the full timeout (Table~\ref{tab:online_outcomes}). Because the calibration-free controllers failed slightly less often, this penalty fell more heavily on the calibrated group, and the strongest calibration-free controllers reached penalized movement times comparable to those of the calibrated controllers. Among the calibration-free controllers, Triplet and RestLoss had the lowest penalized movement time (4.96~s), 0.29~s faster than Base, and RunNorm and Segmented showed similar gains. Throughput followed the same pattern, with the calibrated advantage narrowing from 0.085~b/s for nominal throughput to 0.048~b/s once failures were included. Calibration was therefore associated with faster successful acquisitions, but the size of that advantage depended on whether failures were counted.

\subsection{Trajectory Quality}
\label{sec:trajectory_quality}

\begin{figure}[b]
\centering
\includegraphics[width=0.9\linewidth]{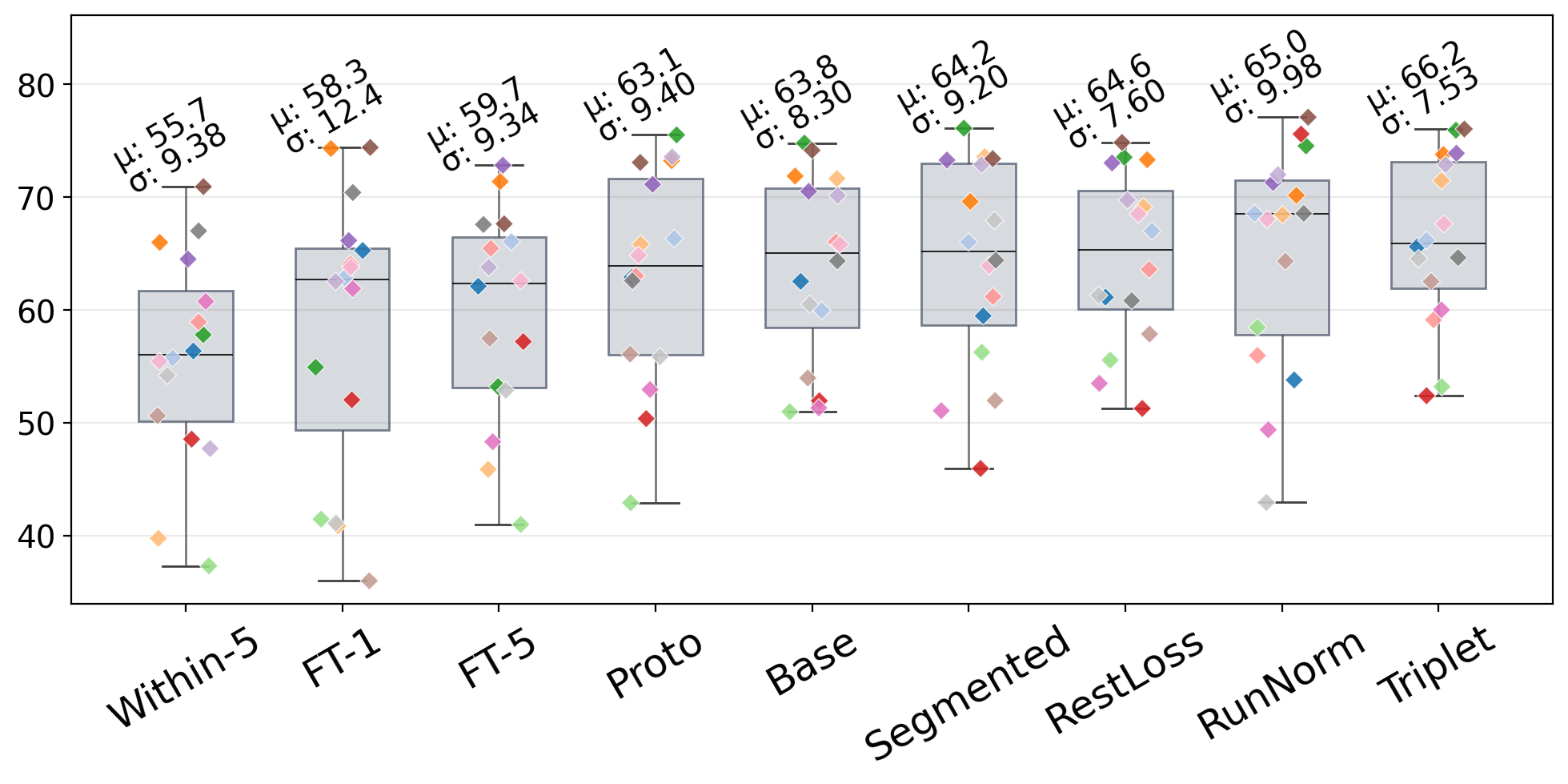}
\caption{Path efficiency of each controller. Every calibration-free controller produced more direct trajectories than every calibrated controller, despite requiring no user-specific training. The other three trajectory-quality measures followed a similar pattern and are reported in Table~\ref{tab:trajectory_quality}.}
\label{fig:trajectory_quality}
\end{figure}

The clearest difference between paradigms was in trajectory quality (Table~\ref{tab:trajectory_quality} and Fig.~\ref{fig:trajectory_quality}). On all four measures, the calibration-free controllers produced more direct and stable cursor movements than the calibrated controllers, with path efficiency of 63.1\% to 66.2\% against 55.7\% to 59.7\%, as well as fewer direction changes, fewer overshoots, and shorter stopping distances. This separation was complete, in that every calibration-free controller outperformed every calibrated controller on every measure, and it persisted after adjusting for cursor speed (Section~\ref{sec:paradigm_comparison}).

\begin{figure}[t]
\centering
\includegraphics[width=\linewidth]{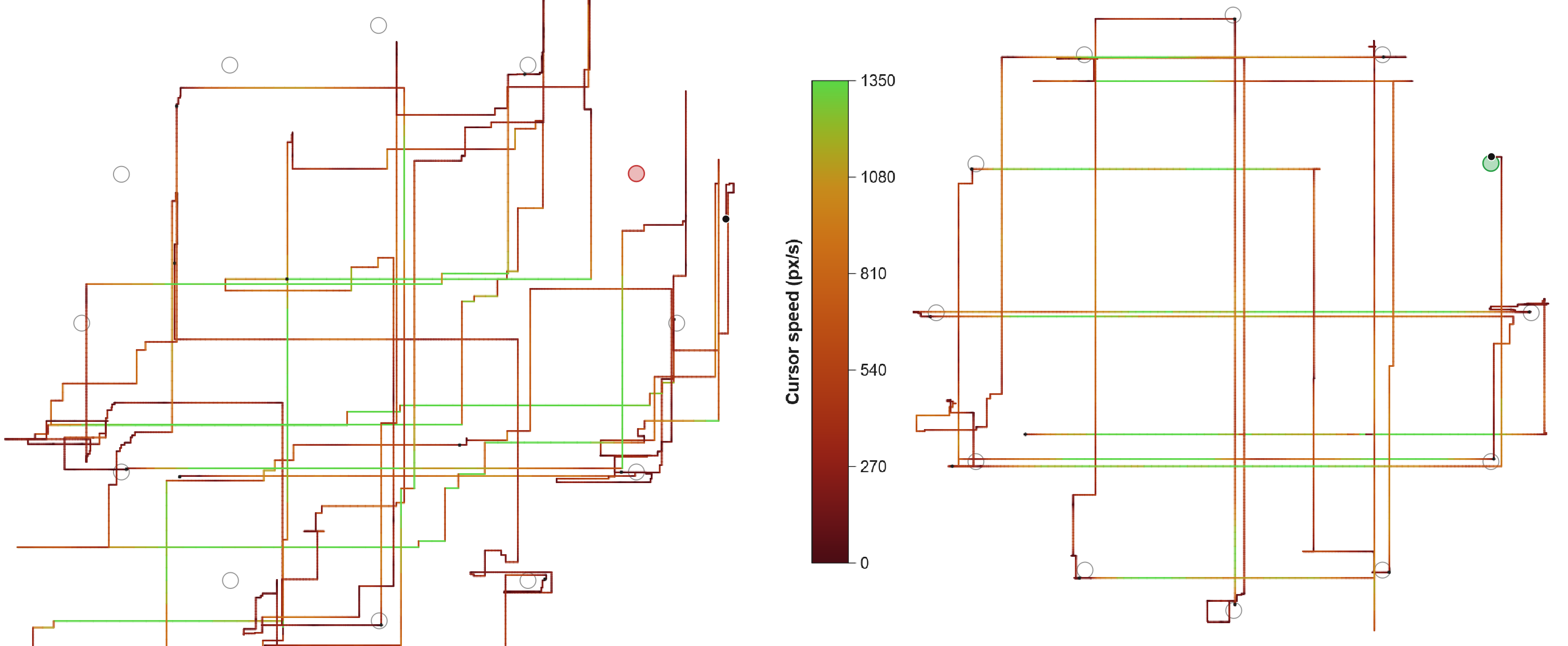}
\caption{Trajectories are from the median participant (rank 9 of 16) when participants were ranked by mean completion rate across all nine models. FT-5 (left) had the highest offline accuracy of the calibrated controllers, and Segmented (right) had the highest completion rate of the calibration-free controllers.}
\label{fig:example_trajectories}
\end{figure}

Within the calibration-free group, Triplet had the highest path efficiency (66.2\%) and the lowest direction-change ratio (2.70), while RunNorm had the fewest overshoots (0.37) and the shortest stopping distance (6.1~px). Segmented improved on Base on all four measures, and RestLoss on all but the direction-change ratio. Stopping distances for the calibrated controllers (19.0 to 21.1~px) were roughly three times those of the best calibration-free controllers (6.1 to 6.7~px). Representative trajectories from FT-5, the most accurate calibrated controller, and Segmented, the calibration-free controller with the highest completion rate, are shown in Fig.~\ref{fig:example_trajectories}.

\begin{table}[t]
\centering
\caption{Trajectory-quality metrics. Higher path efficiency is better, while lower values are preferred for the remaining metrics.}
\label{tab:trajectory_quality}
\small
\begin{tabular}{lcccc}
\toprule
\textbf{Model} & \textbf{Path Eff.} & \textbf{Dir. Ratio} & \textbf{Over.} & \textbf{Stop Dist.} \\
\midrule
Base      & 63.8 & 2.99 & 0.57 & 10.3 \\
Proto     & 63.1 & 3.17 & 0.53 & 11.6 \\
RestLoss  & 64.6 & 3.03 & 0.42 &  6.7 \\
Triplet   & 66.2 & 2.70 & 0.44 &  7.7 \\
RunNorm   & 65.0 & 2.86 & 0.37 &  6.1 \\
Segmented & 64.2 & 2.90 & 0.44 &  9.0 \\
\midrule
Within-5  & 55.7 & 4.74 & 0.74 & 19.0 \\
FT-1      & 58.3 & 3.61 & 0.68 & 21.1 \\
FT-5      & 59.7 & 3.30 & 0.65 & 19.5 \\
\bottomrule
\end{tabular}

\end{table}

\subsection{Intervention-Level Comparisons}
\label{sec:family_a}

Each calibration-free intervention was compared with Base using the Family~A contrasts (Table~\ref{tab:familyA}). All five improved completion rate, by 3.0 to 7.6 percentage points, and all five reduced penalized movement time. Four of the five also improved the trajectory-quality composite, the exception being Proto.

These effects were consistent in direction, with rank-biserial correlations for completion rate between 0.22 and 0.71, but only Triplet's reduction in penalized movement time survived Holm--Bonferroni correction ($r_{rb}=-0.79$, Holm $p=0.017$). With sixteen participants and five contrasts per outcome, the family has limited power for effects of this size. These results therefore indicate the direction and approximate size of each intervention's effect, rather than showing that any single intervention is significantly better than Base.

\begin{table*}[t]
\centering
\caption{Family~A contrasts comparing each calibration-free intervention against Base across the sixteen participants. Positive differences favour the intervention for completion rate and the quality composite; negative differences favour it for penalized movement time. $r_{rb}$: matched-pairs rank-biserial correlation. $p$-values are Holm--Bonferroni corrected within each outcome.}
\label{tab:familyA}
\footnotesize
\setlength{\tabcolsep}{5pt}
\begin{tabular}{lccc ccc ccc}
\toprule
& \multicolumn{3}{c}{Completion Rate}
& \multicolumn{3}{c}{Quality Composite}
& \multicolumn{3}{c}{Pen. Movement Time} \\
\cmidrule(lr){2-4}\cmidrule(lr){5-7}\cmidrule(lr){8-10}
Intervention & Diff. & $r_{rb}$ & $p$ & Diff. & $r_{rb}$ & $p$ & Diff. & $r_{rb}$ & $p$ \\
\midrule
Segmented & $+7.6$ & $0.59$ & $0.18$  & $+0.17$ & $0.47$  & $0.31$  & $-0.19$ & $-0.37$ & $0.63$ \\
Triplet   & $+5.9$ & $0.71$ & $0.078$ & $+0.28$ & $0.60$  & $0.13$  & $-0.29$ & $-0.79$ & $0.017$ \\
RestLoss  & $+5.7$ & $0.68$ & $0.10$  & $+0.23$ & $0.68$  & $0.078$ & $-0.29$ & $-0.57$ & $0.18$ \\
RunNorm   & $+4.3$ & $0.22$ & $0.46$  & $+0.33$ & $0.47$  & $0.31$  & $-0.20$ & $-0.37$ & $0.63$ \\
Proto     & $+3.0$ & $0.37$ & $0.44$  & $-0.04$ & $-0.25$ & $0.40$  & $-0.09$ & $-0.15$ & $0.63$ \\
\bottomrule
\end{tabular}
\end{table*}

\subsection{Paradigm-Level Comparison}
\label{sec:paradigm_comparison}

Pooling controllers by paradigm (Family~B, Table~\ref{tab:paradigm_comparison}) showed no significant difference in completion rate, which was 1.8 percentage points higher for calibration-free control ($p=0.76$), or penalized movement time ($p=0.27$). Practical time to completion was therefore comparable between paradigms despite the offline accuracy gap.

The largest effect in the study was in trajectory quality. The calibration-free paradigm scored higher on the quality composite in every participant ($r_{rb}=1.00$, $d_z=1.33$, $p<0.001$), and this was the only Family~B contrast to survive correction.

The calibrated paradigm was faster. Movement time on successful acquisitions was 0.15~s shorter ($p=0.018$) and penalized nominal throughput was 0.05~b/s higher ($p=0.034$), whereas effective throughput differed by only 0.02~b/s ($p=0.27$). Neither speed difference survived Holm--Bonferroni correction (Holm $p=0.090$ and $0.136$), so the calibrated controllers were faster when they succeeded, but this did not make them more reliable.

\begin{figure}[t]
\centering
\includegraphics[width=\columnwidth]{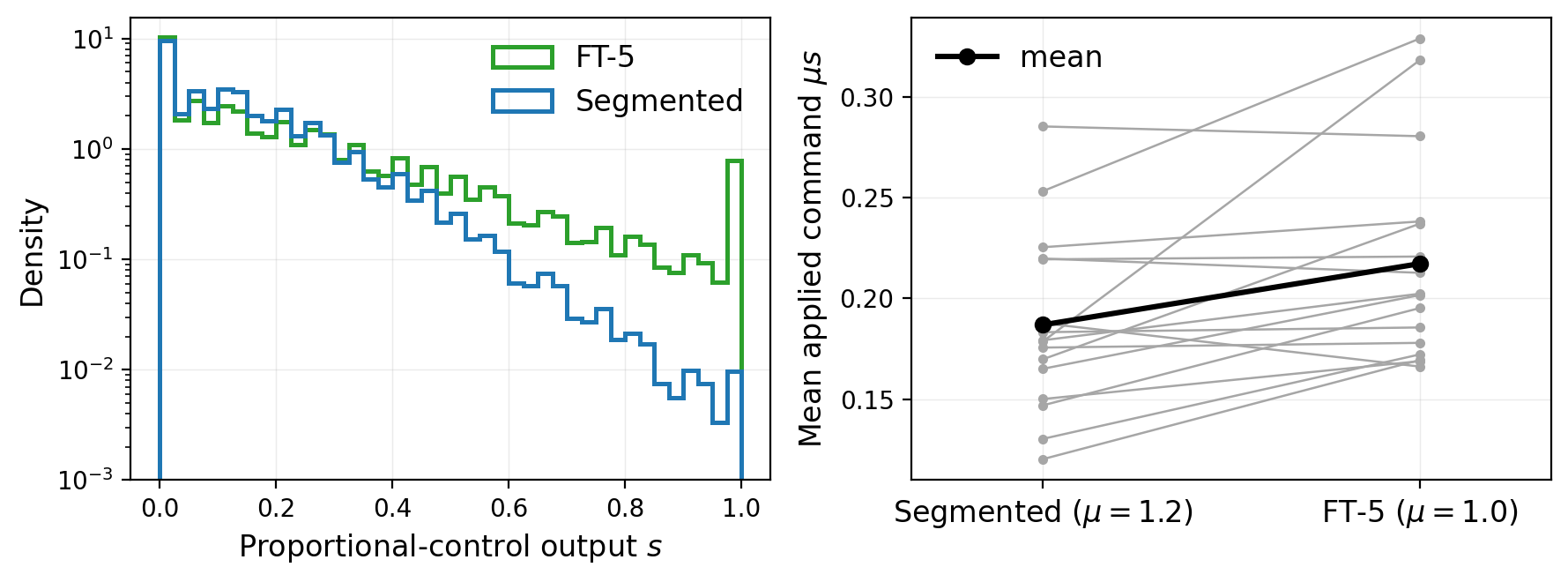}
\caption{Proportional-control speed commands for FT-5 and Segmented, over all frames on which an active class was predicted. Left: distribution of the speed output $s$ before the paradigm velocity multiplier was applied, shown as a histogram normalized to unit area with a logarithmic vertical axis. FT-5 produced high outputs more often and reached full speed ($s=1$) on 2.4\% of frames, against 0.04\% for Segmented. Right: mean speed command after the multiplier ($\mu \cdot s$, as a fraction of full speed) for each participant, with grey lines joining each participant's values for the two controllers and the group mean in black. The multiplier applied to the calibration-free controllers ($\mu=1.2$) raised Segmented's mean command from 0.156 to 0.187, halving the gap to FT-5 (0.217) without closing it.}
\label{fig:velocity_command}
\end{figure}

Part of the speed difference is attributable to the control mapping rather than to the classifier. Applied to each participant's own screen-guided repetitions, the thresholds derived from the pooled training data set $\theta^k_{hi}$ on average 2.21 times higher than the individually fitted value, ranging from 1.95 for wrist extension to 2.41 for hand open. On those same contractions, the individual mapping produced a mean output of $s=0.50$ and reached full speed on 20.6\% of windows, against $s=0.21$ and 0.3\% for the pooled-data mapping. Equalizing the two would have required a median multiplier of 2.19 (interquartile range 1.85 to 3.07), so the applied value of 1.2 under-corrected the pooled-data mapping for every participant. The same pattern held during the task (Fig.~\ref{fig:velocity_command}). This comparison used Segmented, the calibration-free controller with the highest completion rate. All calibration-free controllers shared the same pooled-data mapping and velocity multiplier, so the comparison reflects their shared speed mapping rather than the choice of classifier. Over all frames on which an active class was predicted, FT-5 produced a mean output of $s=0.217$ against 0.156 for Segmented and reached full speed on 2.4\% of frames against 0.04\%, with every participant ordered the same way ($r_{rb}=-1.00$, Holm $p<0.001$ for both). The multiplier narrowed this gap but did not close it, leaving applied commands of 0.187 against 0.217 ($r_{rb}=-0.75$, Holm $p=0.013$).

Because slower cursor movement can by itself reduce overshoot and stopping error, the paradigm contrast on each trajectory measure was repeated in a mixed model with trial speed, the mean on-screen cursor speed over frames in which the cursor moved, as a covariate and participant and controller as random effects. The trial speeds of the two paradigms largely overlapped, so the comparison did not rely on extrapolation. Adjusting for speed left the paradigm gap essentially unchanged for path efficiency ($+6.6$ to $+7.0$ percentage points) and stopping distance ($-9.5$~px in both cases), the 95\% confidence interval excluded zero (one for the overshoot rate ratio) for all four measures, and the gap widened at higher trial speeds for all but stopping distance. Normalizing stopping distance by the mean cursor speed over the 150~ms before target entry gave the same result, with stopping distance equivalent to 0.05~s of continued travel at entry speed under calibration-free control against 0.15~s under calibrated control ($p=0.008$, Wilcoxon signed-rank). At matched entry speed, calibration-free control had lower odds of overshooting (odds ratio 0.70, 95\% confidence interval 0.64 to 0.76, logistic mixed model with entry speed as a covariate). Part of the speed difference may also reflect participants slowing down for a controller they trusted less, in which case adjusting for speed may remove part of a real effect.

Because Family~B pools all six calibration-free controllers, including the weaker Base and Proto, Family~C compared the strongest controller of each paradigm directly. Applying the selection rule of Section~\ref{sec:stats} to the offline results in Table~\ref{tab:offline_results} identified Triplet and FT-5 as the strongest controllers. Because Segmented was within 0.4 percentage points of Triplet offline, it was also compared against FT-5 as a secondary contrast. The trajectory-quality advantage held under this stricter comparison, with Triplet scoring 0.73 higher than FT-5 on the composite ($r_{rb}=0.91$, $d_z=1.26$, Holm $p=0.003$). No other outcome differed significantly. Triplet completed 3.8 percentage points more targets ($p=0.37$), FT-5 was 0.11~s faster on successful acquisitions ($p=0.16$), and the two were within 0.01~s in penalized movement time ($p=0.43$), with differences in effective and penalized nominal throughput of 0.04 and 0.06~b/s favouring FT-5 ($p=0.23$ and $p=0.10$, all Holm $p\geq0.52$). The secondary contrast, Segmented against FT-5, reproduced the quality result at a similar magnitude (0.62, $r_{rb}=0.87$, $d_z=1.00$, Holm $p=0.005$). Segmented completed 5.5 percentage points more targets ($p=0.61$), but FT-5 retained a 0.43~s advantage in movement time on successful acquisitions ($r_{rb}=0.88$, $d_z=0.96$, Holm $p=0.005$), which did not carry over to penalized movement time (0.09~s, $p=0.35$) or to either throughput measure after correction.

\begin{table}[t]
\centering
\caption{Calibration-free versus calibrated control across the sixteen participants. Differences are calibration-free minus calibrated. Higher is better for completion rate, quality composite, and throughput; lower is better for movement time. $r_{rb}$: matched-pairs rank-biserial correlation; $d_z$: Cohen's $d_z$. Holm $p$ applies the Holm--Bonferroni correction across the six contrasts of Family~B. For completion rate, the positive Diff. and negative $r_{rb}$ reflect right-skewed paired differences.}
\label{tab:paradigm_comparison}
\small
\setlength{\tabcolsep}{4pt}
\begin{tabular}{lccccc}
\toprule
\textbf{Metric} & \textbf{Diff.} & \textbf{$r_{rb}$} & \textbf{$d_z$} & \textbf{$p$} & \textbf{Holm $p$} \\
\midrule
Quality Composite          & $+0.79$ & $1.00$  & $1.33$  & $<0.001$ & $<0.01$ \\
Completion Rate (\%)       & $+1.8$  & $-0.09$ & $0.10$  & $0.76$   & $0.81$ \\
Movement Time (s)          & $+0.15$ & $0.66$  & $0.62$  & $0.018$  & $0.090$ \\
MT Pen. (s)                & $+0.06$ & $0.32$  & $0.10$  & $0.27$   & $0.81$ \\
Throughput Eff. (b/s)      & $-0.02$ & $-0.32$ & $-0.22$ & $0.27$   & $0.81$ \\
\shortstack[l]{Throughput\\Nom. Pen. (b/s)} & $-0.05$ & $-0.60$ & $-0.43$ & $0.034$  & $0.136$ \\
\bottomrule
\end{tabular}

\end{table} 

\subsection{Offline Accuracy and the Benefit of Calibration}
\label{sec:offline_online}

\begin{figure*}[t]
\centering
\includegraphics[width=0.8\textwidth]{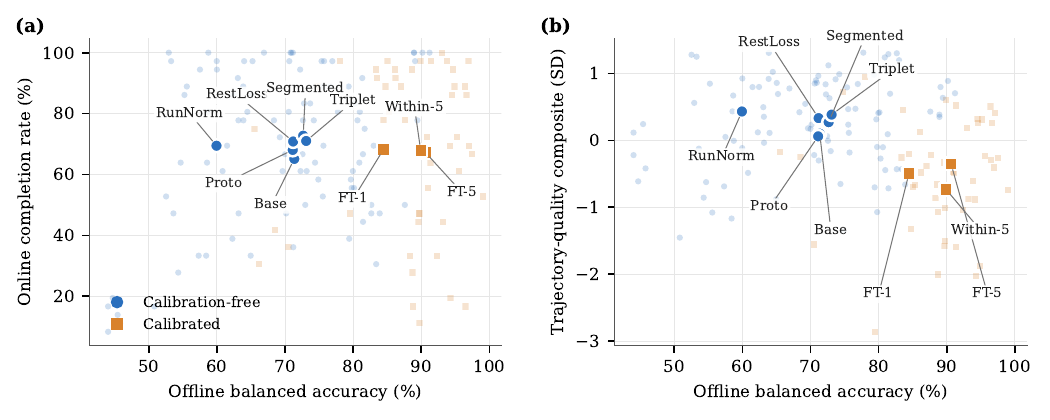}
\caption{Offline balanced accuracy against (a) online completion rate and (b) the trajectory-quality composite. Small markers are individual participant-by-controller cells, and large labelled markers are controller means. All controllers, including Segmented, were scored offline on the same unsegmented held-out repetition. Offline accuracy did not predict completion rate, and across paradigms the more accurate calibrated controllers produced poorer trajectories.}
\label{fig:offline_online}
\end{figure*}

Offline balanced accuracy did not predict completion rate at any level of analysis (Fig.~\ref{fig:offline_online}a). Across the nine controller means, the Spearman correlation was $\rho=-0.27$ ($p=0.49$), and across the six calibration-free controllers it was $\rho=+0.54$ ($p=0.27$). Neither relationship held within individual participants. Correlating the two measures across controllers separately for each participant gave a median $\rho$ of $+0.08$ for all nine controllers and $-0.03$ for the calibration-free controllers (Wilcoxon signed-rank against zero, $p=0.98$ and $p=0.84$). Participants who were more accurate offline were also not more reliable online ($\rho=+0.16$, $p=0.56$, using participant means across controllers), and no controller showed a significant correlation across participants (all $|\rho|\leq0.29$). The offline range among the calibration-free controllers was narrow, however, with five of the six within 2 percentage points of one another, which limits the power of the within-paradigm comparison.

Trajectory quality showed an inverse relationship across paradigms (Fig.~\ref{fig:offline_online}b). Within each participant, controllers with higher offline accuracy produced lower trajectory quality (median $\rho=-0.44$, negative in 15 of 16 participants, $p<0.001$) and longer stopping distances in all 16 participants, whereas no relationship was present among the calibration-free controllers (median $\rho=0.00$, $p=0.64$). A mixed model with participant as a random intercept showed that this inverse relationship came from the paradigm rather than from accuracy itself. A within-participant increase of 10 percentage points in offline accuracy was associated with a 0.25~SD reduction in the quality composite ($p<0.001$), but this effect disappeared once paradigm was added to the model (0.02~SD, $p=0.74$), and calibrated controllers then scored 0.83~SD lower than calibration-free controllers of equal offline accuracy ($p<0.001$).

\section{Discussion}
\label{sec:discussion}

\subsection{Why Calibration-Free Control Performed Well}
\label{sec:discussion_why}

Offline, FT-5 was between 17.5 and 30.7 percentage points more accurate than the calibration-free controllers, depending on the comparison. This advantage failed to carry over to closed-loop control in a direct way, though. Completion rate did not differ between paradigms, trajectory quality strongly favoured calibration-free control, and although the calibrated controllers were faster on successful acquisitions, it was not significantly after correction and partly because of their individually fitted speed mapping. The two paradigms therefore differed along separate dimensions of performance, and the offline accuracy gap did not predict any of them well.

The trajectory-quality result yielded both the largest effect in the study and the opposite of what offline accuracy would predict. It could not be explained by slower movement, because the advantage persisted at matched cursor speeds and at matched entry speeds (Section~\ref{sec:paradigm_comparison}). It was also not a product of the interventions, because even Base outperformed every calibrated controller on all four trajectory measures (Table~\ref{tab:trajectory_quality}). Nor was it explained by the difference in offline accuracy, because at matched offline accuracy the calibrated controllers still scored 0.83~SD lower on the trajectory-quality composite (Section~\ref{sec:offline_online}). The two measures least dependent on speed by definition, the direction-change ratio and stopping behaviour at a given entry speed, instead point to classifier outputs that remained more stable during closed-loop use, particularly around the transitions between movement and rest.

One explanation is that calibration improves accuracy on a different kind of data than closed-loop control produces. A calibration session samples the user's behaviour over a few minutes, in a single posture and with the contraction style prompted by the screen. A model fitted closely to that sample classifies data from the same session well, which offline accuracy registers, but it may remain sensitive to the changes in effort, posture, and strategy that appear once the user is pursuing a target, since screen-guided training is known not to capture goal-oriented behaviour~\cite{campbell2024screen}. A model trained across hundreds of users cannot fit any one session's idiosyncrasies, so it is pushed toward features that hold across people, and these may be the same features that remain stable when a user's own behaviour shifts. This is consistent with the broader observation in machine learning that training across heterogeneous data favours representations that are robust to distribution shift~\cite{ganin2016dann,levy2020large}. 

The fine-tuned controllers illustrate this point. FT-5 differed from Base only in being fine-tuned on five calibration repetitions from the participant. This raised offline accuracy by 19.3 percentage points, yet FT-5 produced poorer trajectories than Base on all four measures (Table~\ref{tab:trajectory_quality}). Part of this difference may come from the individually fitted speed mapping. However, fine-tuning all layers of a pretrained network has been shown to distort its learned features and reduce robustness to distribution shift, even as accuracy on data resembling the fine-tuning set improves~\cite{kumar2022finetuning}. A similar effect has been reported for EMG, where fully fine-tuning a large generic decoder overfit it to the target user and occasionally degraded performance for users the generic model already served well~\cite{ctrl2024generic}. These findings are consistent with calibration reducing robustness to the changes in behaviour that occur during closed-loop use. Future work should directly test this by comparing the stability of each controller's output on screen-guided and goal-directed data, and by varying how much of the network is fine-tuned.

If correct, this account also changes how inter-user variability should be viewed. Differences in electrode placement, limb position, effort, and skin impedance are usually treated as nuisance factors that recalibration exists to correct~\cite{campbell2020current}. The present results suggest that at least part of this variability is structured and shared across users, so that a sufficiently diverse population can teach a model to tolerate it directly. Recalibration is then one way of handling variability rather than the only one, and which sources of variability are better handled by training diversity and which genuinely require data from the individual should be considered an ongoing question.

\subsection{Implications for Deployment}
\label{sec:discussion_deployment}

The results of this work highlight that the benefit of calibration should no longer be assumed when large amounts of training data are available for matched hardware. This is consistent with the generic decoder reported by CTRL-labs, which supported closed-loop control without per-user calibration~\cite{ctrl2024generic}. The further 16\% gain they reported from personalization was measured in decoding performance, which is comparable in nature to classification accuracy here, so it did not establish a closed-loop benefit. In the present study, a comparable gain in offline accuracy did not produce more reliable closed-loop control. The calibrated controllers, as evaluated, retained an advantage in nominal speed, and there may be applications in which that advantage matters. Even so, part of this advantage came from the speed mapping rather than the classifier, because the speed thresholds of the calibration-free controllers were derived from the pooled training data rather than from each user's own contractions, and the velocity multiplier only partly offset this (Section~\ref{sec:paradigm_comparison}). Since their trajectory-quality advantage held at matched speeds, a better-tuned mapping could allow them to move faster without losing it. One way to achieve this without giving up calibration-free operation would be to adapt the speed mapping during use, in the same way that RunNorm adapts the input statistics, starting from the population range and converging toward the user's own.

This could matter most for consumer and general-purpose HCI applications, where calibration would be a fixed cost imposed on every new user before the system can be used at all~\cite{saponas2009enable,ctrl2024generic}. Where stable trajectories and immediate usability matter more than small gains in movement time, skipping calibration may be the better default rather than an acceptable compromise.

A shared cross-user model also changes how a myoelectric system can improve over time. Within-user models are rebuilt for every user and often every session, so what is learned from one deployment carries over to the next only as general design guidance. A single population model, by contrast, can be refined directly, for example by adding training users, correcting problematic training data, or targeting the users it serves poorly, and every subsequent user benefits from each change. The interventions evaluated here illustrate this on a small scale, since each was applied to the same backbone and could be combined and carried forward to all future users without any additional data collection.

More broadly, this work suggests that calibration should be justified by whether it improves closed-loop performance, rather than by offline accuracy alone.

\subsection{Offline Accuracy and Closed-Loop Performance}
\label{sec:discussion_offline_online}

The disconnect between offline and online performance has been documented for decades in calibrated systems~\cite{Lock2005Realtime,Hargrove2007Realtime,ortizcatalan2015offline,lu2019offline}. The present results show that it also holds between calibrated and calibration-free control, and here the relationship was inverted; the paradigm with lower offline accuracy produced the more stable closed-loop trajectories. A similar inversion was reported for an adaptive scheme that began without user data, which outperformed a screen-guided baseline in closed loop despite lower offline accuracy~\cite{campbell2024screen}. Part of the reason may be that offline accuracy treats every misclassification of prerecorded data identically, whereas during closed-loop use the user responds to the system's behaviour, and the same error rate can have very different consequences depending on when the errors occur~\cite{campbell2024CIIL}.

The present results also go beyond the usual observation that offline accuracy is a noisy predictor of online performance. Offline accuracy did not predict completion rate across controllers, within participants, or across participants, and across paradigms it was inversely related to trajectory quality, an association explained by calibration rather than by accuracy itself. Calibration therefore had the opposite effect to that intended, because the controllers it made more accurate offline produced poorer closed-loop trajectories. Within the calibration-free group, the offline differences were too small to test whether accuracy would track closed-loop performance among classifiers that share a control framework, so the results do not rule this out. They do show, however, that offline accuracy cannot be relied on to choose between calibrated and calibration-free control. Offline measures beyond classification accuracy, such as those describing the separability and repeatability of the training data, have predicted online usability better than accuracy in calibrated systems~\cite{Nawfel2021Multivariate,Lv2019}, and whether such measures can capture the trajectory differences observed here is a direction for future work.

RunNorm illustrated a narrower problem. Its weak offline score came from the structure and length of the test recording rather than from its representation, because a class-blocked repetition supplies normalization statistics that no real stream of use would produce. Offline protocols designed for static classifiers may therefore misrepresent controllers that adapt during use, and such controllers are better evaluated online or with test recordings that resemble continuous use.

\subsection{Limitations}
\label{sec:discussion_limitations}

Sixteen participants were sufficient to detect the trajectory-quality effect, which held both when paradigms were pooled and when only the strongest controller of each paradigm was compared. They were not sufficient for most of the intervention-level effects in Table~\ref{tab:familyA}, only one of which survived correction, or for the speed contrasts in Table~\ref{tab:paradigm_comparison}, so a larger cohort would be needed to fully establish these statistically. The offline--online correspondence analyses were exploratory, and the narrow range of offline accuracies across the calibration-free controllers likely limited the within-paradigm comparison. Also, because the sixth repetition used to score offline accuracy was also used for early stopping of the calibrated controllers, their offline accuracies are likely optimistic, which if anything overstates the offline advantage of calibration.

All experiments were conducted within a single session, with the calibrated controllers trained immediately before use. This is the most favourable condition for calibration, because calibrated models degrade across days, after donning and doffing, and with electrode shift~\cite{hwang2017real,vidovic2016improving}, whereas the calibration-free controllers were never fitted to a particular session and treat each new session as they would a new user. A multi-session study would therefore be expected to favour calibration-free control further rather than less, although this was not tested here.

The two paradigms were compared as complete control systems rather than as isolated classifiers, because the speed thresholds and the velocity multiplier differed along with the classifier. The velocity multiplier (Section~\ref{sec:cursor_control}) was applied only to the calibration-free controllers and was chosen through informal piloting. It under-corrected the pooled-data speed mapping, and calibration-free control rarely reached the upper end of the proportional range, commanding full speed on 0.04\% of active frames against 2.4\% for FT-5 (Fig.~\ref{fig:velocity_command}). The speed comparisons between paradigms therefore reflect the control mapping as well as the classifier.

All calibration-free controllers, and the two fine-tuned calibrated controllers, shared the MHCNN backbone. Holding the architecture fixed allowed differences between controllers to be attributed to the training or preprocessing intervention, but it also means that the current calibration-free results can only strictly be considered as representative of this one architecture rather than calibration-free control in general. Because the MHCNN is a small network (89,189 parameters) selected from a modest benchmark, larger or more expressive architectures may well improve on it, so these results are likely a lower bound on what calibration-free control can achieve than its ceiling.

Similarly, the cross-user controllers were trained on only the 306 training users of EMG-EPN-612, because the remaining users were held out for offline model development and selection. Eddy \emph{et al.}~\cite{eddy2024big} found that cross-user accuracy continued to improve as the number of training users increased, so retraining on the full dataset may further improve closed-loop performance and should be tested in future work.

The cross-user models were also trained and tested with a single device, electrode configuration, and gesture set. The results should therefore not be taken as evidence that arbitrary EMG datasets can be pooled into a universally deployable controller, because differences in hardware, electrode placement, and task definition introduce domain shifts that this study did not test~\cite{ganin2016dann}.

Finally, all participants were able-bodied. People with limb difference or with neurological conditions such as spinal cord injury or stroke may also benefit from these interfaces, but their EMG can differ substantially from that of the able-bodied population on which these models were trained, so some user-specific adaptation is likely to be needed~\cite{lu2019offline,campbell2020comparison}. 
The behaviour of the fine-tuned controllers (Section~\ref{sec:discussion_why}) suggests that how this adaptation is done matters, and future work should examine strategies that better preserve the stability of the cross-user model, such as freezing the early layers~\cite{hajian2024generalizing} or limiting how far fine-tuning moves the model from its population-trained weights.

\section{Conclusion}
\label{sec:conclusion}

This study asked whether user-specific calibration is still necessary when a myoelectric controller is trained across hundreds of users. Six calibration-free controllers and three calibrated controllers were compared under a common closed-loop Fitts' law protocol. Although the calibrated controllers were more accurate offline, they were not more reliable online, and the calibration-free controllers achieved comparable completion rates without any training data from the user. Offline accuracy did not predict closed-loop reliability, and across paradigms, the additional accuracy provided by calibration was associated with poorer trajectories in nearly every participant.

The clearest difference between the paradigms was in trajectory quality. The calibration-free controllers produced more direct trajectories, fewer corrections and overshoots, and shorter stopping distances than every calibrated controller, and this advantage held in every participant and after adjusting for cursor speed. The calibrated controllers were faster on successful acquisitions, but this difference did not survive correction for multiple comparisons and was partly due to their individually fitted speed mapping rather than their classifiers.

Within the conditions tested here, user-specific calibration was therefore not necessary, and the offline accuracy it provided did not translate into better closed-loop control. As larger cross-user datasets become available, calibration-free models offer a practical route toward plug-and-play myoelectric interfaces for both assistive and general human-computer interaction. Future work should treat the role of calibration as an open question, asking how to make it more efficient and under what conditions it is needed at all.

\section*{Acknowledgment}

Claude (Opus 5.5) was used to support the research and writing of this manuscript. All generated material was critically reviewed, revised, and verified by the authors.

\bibliographystyle{IEEEtran}
\bibliography{mybibliography}

\begin{thebibliography}{10}
\providecommand{\url}[1]{#1}
\csname url@samestyle\endcsname
\providecommand{\newblock}{\relax}
\providecommand{\bibinfo}[2]{#2}
\providecommand{\BIBentrySTDinterwordspacing}{\spaceskip=0pt\relax}
\providecommand{\BIBentryALTinterwordstretchfactor}{4}
\providecommand{\BIBentryALTinterwordspacing}{\spaceskip=\fontdimen2\font plus
\BIBentryALTinterwordstretchfactor\fontdimen3\font minus \fontdimen4\font\relax}
\providecommand{\BIBforeignlanguage}[2]{{%
\expandafter\ifx\csname l@#1\endcsname\relax
\typeout{** WARNING: IEEEtran.bst: No hyphenation pattern has been}%
\typeout{** loaded for the language `#1'. Using the pattern for}%
\typeout{** the default language instead.}%
\else
\language=\csname l@#1\endcsname
\fi
#2}}
\providecommand{\BIBdecl}{\relax}
\BIBdecl

\bibitem{scheme2011electro}
E.~Scheme and K.~Englehart, ``\BIBforeignlanguage{en}{Electromyogram pattern recognition for control of powered upper-limb prostheses: {State} of the art and challenges for clinical use},'' \emph{\BIBforeignlanguage{en}{The Journal of Rehabilitation Research and Development}}, vol.~48, no.~6, p. 643, 2011.

\bibitem{englehart2003robust}
K.~Englehart and B.~Hudgins, ``\BIBforeignlanguage{en}{A robust, real-time control scheme for multifunction myoelectric control},'' \emph{\BIBforeignlanguage{en}{IEEE Transactions on Biomedical Engineering}}, vol.~50, no.~7, pp. 848--854, Jul. 2003.

\bibitem{saponas2008demonstrate}
T.~S. Saponas, D.~S. Tan, D.~Morris, and R.~Balakrishnan, ``Demonstrating the feasibility of using forearm electromyography for muscle-computer interfaces,'' in \emph{Proceedings of the SIGCHI Conference on Human Factors in Computing Systems}, ser. CHI '08.\hskip 1em plus 0.5em minus 0.4em\relax New York, NY, USA: Association for Computing Machinery, 2008, p. 515–524.

\bibitem{saponas2009enable}
T.~S. Saponas, D.~S. Tan, D.~Morris, R.~Balakrishnan, J.~Turner, and J.~A. Landay, ``Enabling always-available input with muscle-computer interfaces,'' p. 167–176, 2009.

\bibitem{ctrl2024generic}
P.~Kaifosh, T.~R. Reardon, {CTRL-labs at Reality Labs} \emph{et~al.}, ``\BIBforeignlanguage{en}{A generic non-invasive neuromotor interface for human-computer interaction},'' \emph{\BIBforeignlanguage{en}{Nature}}, vol. 645, no. 8081, pp. 702--711, Sep. 2025.

\bibitem{scheme2013fitts}
E.~J. Scheme and K.~B. Englehart, ``Validation of a selective ensemble-based classification scheme for myoelectric control using a three-dimensional fitts' law test,'' \emph{IEEE Transactions on Neural Systems and Rehabilitation Engineering}, vol.~21, no.~4, pp. 616--623, 2013.

\bibitem{hahne2014linear}
J.~M. Hahne, F.~Bie{\ss}mann, N.~Jiang, H.~Rehbaum, D.~Farina, F.~C. Meinecke, K.-R. M{\"u}ller, and L.~C. Parra, ``Linear and nonlinear regression techniques for simultaneous and proportional myoelectric control,'' \emph{IEEE Transactions on Neural Systems and Rehabilitation Engineering}, vol.~22, no.~2, pp. 269--279, 2014.

\bibitem{jiang2014intuitive}
N.~Jiang, H.~Rehbaum, I.~Vujaklija, B.~Graimann, and D.~Farina, ``Intuitive, online, simultaneous, and proportional myoelectric control over two degrees-of-freedom in upper limb amputees,'' \emph{IEEE Transactions on Neural Systems and Rehabilitation Engineering}, vol.~22, no.~3, pp. 501--510, 2014.

\bibitem{campbell2021deep}
E.~Campbell, A.~Phinyomark, and E.~Scheme, ``Deep {Cross}-{User} {Models} {Reduce} the {Training} {Burden} in {Myoelectric} {Control},'' \emph{Frontiers in Neuroscience}, vol.~15, p. 657958, May 2021.

\bibitem{hwang2017real}
H.-J. Hwang, J.~M. Hahne, and K.-R. M{\"u}ller, ``Real-time robustness evaluation of regression based myoelectric control against arm position change and donning/doffing,'' \emph{PLoS ONE}, vol.~12, no.~11, p. e0186318, 2017.

\bibitem{vidovic2016improving}
M.~M.-C. Vidovic, H.-J. Hwang, S.~Ams{\"u}ss, J.~M. Hahne, D.~Farina, and K.-R. M{\"u}ller, ``Improving the robustness of myoelectric pattern recognition for upper limb prostheses by covariate shift adaptation,'' \emph{IEEE Transactions on Neural Systems and Rehabilitation Engineering}, vol.~24, no.~9, pp. 961--970, Sep. 2016.

\bibitem{biddiss2007upper}
E.~A. Biddiss and T.~T. Chau, ``Upper limb prosthesis use and abandonment: A survey of the last 25 years,'' \emph{Prosthetics and Orthotics International}, vol.~31, no.~3, pp. 236--257, 2007.

\bibitem{salminger2022current}
S.~Salminger, H.~Stino, L.~H. Pichler, C.~Gstoettner, A.~Sturma, J.~A. Mayer, M.~Szivak, and O.~C. Aszmann, ``Current rates of prosthetic usage in upper-limb amputees --- have innovations had an impact on device acceptance?'' \emph{Disability and Rehabilitation}, vol.~44, no.~14, pp. 3708--3713, 2022.

\bibitem{ulysse2017transfer}
U.~Cote-Allard, C.~L. Fall, A.~Campeau-Lecours, C.~Gosselin, F.~Laviolette, and B.~Gosselin, ``Transfer learning for {sEMG} hand gestures recognition using convolutional neural networks,'' pp. 1663--1668, Oct. 2017.

\bibitem{coteallard2019deep}
U.~C{\^o}t{\'e}-Allard, C.~L. Fall, A.~Drouin, A.~Campeau-Lecours, C.~Gosselin, K.~Glette, F.~Laviolette, and B.~Gosselin, ``Deep learning for electromyographic hand gesture signal classification using transfer learning,'' \emph{IEEE Transactions on Neural Systems and Rehabilitation Engineering}, vol.~27, no.~4, pp. 760--771, 2019.

\bibitem{wu2023transfer}
D.~Wu, J.~Yang, and M.~Sawan, ``Transfer learning on electromyography ({EMG}) tasks: Approaches and beyond,'' \emph{IEEE Transactions on Neural Systems and Rehabilitation Engineering}, vol.~31, pp. 3015--3034, 2023.

\bibitem{li2024across}
W.~Li, X.~Zhang, P.~Shi, S.~Li, P.~Li, and H.~Yu, ``Across sessions and subjects domain adaptation for building robust myoelectric interface,'' \emph{IEEE Transactions on Neural Systems and Rehabilitation Engineering}, vol.~32, pp. 2005--2015, 2024.

\bibitem{eddy2024big}
E.~Eddy, E.~Campbell, S.~Bateman, and E.~Scheme, ``Big data in myoelectric control: large multi-user models enable robust zero-shot {EMG}-based discrete gesture recognition,'' \emph{Frontiers in Bioengineering and Biotechnology}, vol.~12, p. 1463377, Sep. 2024.

\bibitem{xu2024crossuser}
M.~Xu, X.~Chen, Y.~Ruan, and X.~Zhang, ``Cross-user electromyography pattern recognition based on a novel spatial-temporal graph convolutional network,'' \emph{IEEE Transactions on Neural Systems and Rehabilitation Engineering}, vol.~32, pp. 72--82, 2024.

\bibitem{epn}
M.~E. Benalcazar, L.~Barona, L.~Valdivieso, X.~Aguas, and J.~Zea, ``{EMG-EPN-612} dataset,'' Zenodo, 2020.

\bibitem{botros2025zeroshot}
F.~S. Botros, H.~E. Williams, A.~Phinyomark, and E.~J. Scheme, ``From zero- to few-shot: deep temporal learning of wrist {EMG} enables scalable cross-user gesture recognition,'' \emph{Journal of Neural Engineering}, vol.~22, no.~5, p. 056011, 2025.

\bibitem{salter2024emg2pose}
S.~Salter, R.~Warren, C.~Schlager, A.~Spurr, S.~Han, R.~Bhasin, Y.~Cai, P.~Walkington, A.~Bolarinwa, R.~Wang \emph{et~al.}, ``emg2pose: A large and diverse benchmark for surface electromyographic hand pose estimation,'' in \emph{Advances in Neural Information Processing Systems 37 (NeurIPS 2024) Datasets and Benchmarks Track}, 2024, arXiv:2412.02725.

\bibitem{sivakumar2024emg2qwerty}
V.~Sivakumar, J.~Seely, A.~Du, S.~R. Bittner, A.~Berenzweig, A.~Bolarinwa, A.~Gramfort, and M.~I. Mandel, ``emg2qwerty: A large dataset with baselines for touch typing using surface electromyography,'' in \emph{Advances in Neural Information Processing Systems 37 (NeurIPS 2024) Datasets and Benchmarks Track}, 2024, arXiv:2410.20081.

\bibitem{yang2024emgbench}
J.~Yang, M.~Soh, V.~Lieu, D.~J. Weber, and Z.~Erickson, ``{EMGBench}: Benchmarking out-of-distribution generalization and adaptation for electromyography,'' in \emph{Advances in Neural Information Processing Systems 37 (NeurIPS 2024) Datasets and Benchmarks Track}, 2024.

\bibitem{Lock2005Realtime}
B.~A. Lock, K.~Englehart, and B.~Hudgins, ``Real-time myoelectric control in a virtual environment to relate usability vs. accuracy,'' in \emph{Proceedings of the Myoelectric Controls Symposium}, 2005, pp. 17--19.

\bibitem{ortizcatalan2015offline}
M.~Ortiz-Catalan, F.~Rouhani, R.~Br{\aa}nemark, and B.~H{\aa}kansson, ``Offline accuracy: A potentially misleading metric in myoelectric pattern recognition for prosthetic control,'' in \emph{2015 37th Annual International Conference of the IEEE Engineering in Medicine and Biology Society (EMBC)}.\hskip 1em plus 0.5em minus 0.4em\relax IEEE, Aug. 2015, pp. 1140--1143.

\bibitem{Lv2019}
B.~Lv, X.~Sheng, D.~Hao, and X.~Zhu, ``Relationship between offline and online metrics in myoelectric pattern recognition control based on target achievement control test,'' in \emph{Proceedings of the 41st Annual International Conference of the IEEE Engineering in Medicine and Biology Society (EMBC)}, 2019, pp. 6595--6598.

\bibitem{Nawfel2021Multivariate}
J.~L. Nawfel, K.~B. Englehart, and E.~J. Scheme, ``A multi-variate approach to predicting myoelectric control usability,'' \emph{IEEE Transactions on Neural Systems and Rehabilitation Engineering}, vol.~29, pp. 1312--1327, 2021.

\bibitem{jiang2014accurate}
N.~Jiang, I.~Vujaklija, H.~Rehbaum, B.~Graimann, and D.~Farina, ``Is accurate mapping of {EMG} signals on kinematics needed for precise online myoelectric control?'' \emph{IEEE Transactions on Neural Systems and Rehabilitation Engineering}, vol.~22, no.~3, pp. 549--558, 2014.

\bibitem{lu2019offline}
Z.~Lu, A.~Stampas, G.~E. Francisco, and P.~Zhou, ``Offline and online myoelectric pattern recognition analysis and real-time control of a robotic hand after spinal cord injury,'' \emph{Journal of Neural Engineering}, vol.~16, no.~3, p. 036018, Apr. 2019.

\bibitem{campbell2024screen}
E.~Campbell, E.~Eddy, S.~Bateman, B.~Gosselin, U.~C{\^o}t{\'e}-Allard, and E.~Scheme, ``Screen guided training does not capture goal-oriented behaviours: Learning myoelectric control mappings from scratch using context informed incremental learning,'' \emph{IEEE Transactions on Neural Systems and Rehabilitation Engineering}, vol.~33, pp. 332--342, 2024.

\bibitem{simon2011target}
A.~M. Simon, L.~J. Hargrove, B.~A. Lock, and T.~A. Kuiken, ``Target achievement control test: Evaluating real-time myoelectric pattern-recognition control of multifunctional upper-limb prostheses,'' \emph{Journal of Rehabilitation Research and Development}, vol.~48, no.~6, pp. 619--627, 2011.

\bibitem{soukoreff2004towards}
R.~W. Soukoreff and I.~S. MacKenzie, ``\BIBforeignlanguage{en}{Towards a standard for pointing device evaluation, perspectives on 27 years of {Fitts}’ law research in {HCI}},'' \emph{\BIBforeignlanguage{en}{International Journal of Human-Computer Studies}}, vol.~61, no.~6, pp. 751--789, Dec. 2004.

\bibitem{fitts1954information}
P.~M. Fitts, ``The information capacity of the human motor system in controlling the amplitude of movement,'' \emph{Journal of Experimental Psychology}, vol.~47, no.~6, pp. 381--391, 1954.

\bibitem{snell2017prototypical}
J.~Snell, K.~Swersky, and R.~S. Zemel, ``Prototypical networks for few-shot learning,'' 2017.

\bibitem{hermans2017defense}
A.~Hermans, L.~Beyer, and B.~Leibe, ``In defense of the triplet loss for person re-identification,'' 2017.

\bibitem{8516279}
J.~W. Robertson, K.~B. Englehart, and E.~J. Scheme, ``Effects of confidence-based rejection on usability and error in pattern recognition-based myoelectric control,'' \emph{IEEE Journal of Biomedical and Health Informatics}, vol.~23, no.~5, pp. 2002--2008, 2019.

\bibitem{li2025deep}
J.~Li, X.~Jiang, J.~Fan, Y.~Geng, F.~Jia, and C.~Dai, ``Deep end-to-end transfer learning for robust inter-subject and inter-day hand gesture recognition using surface emg,'' \emph{Biomedical Signal Processing and Control}, vol. 100, p. 106892, 2025.

\bibitem{mackenzie1992fitts}
I.~S. MacKenzie, ``\BIBforeignlanguage{en}{Fitts' {Law} as a {Research} and {Design} {Tool} in {Human}-{Computer} {Interaction}},'' \emph{\BIBforeignlanguage{en}{Human–Computer Interaction}}, vol.~7, no.~1, pp. 91--139, Mar. 1992.

\bibitem{scheme2014mnpc}
E.~Scheme, B.~Lock, L.~Hargrove, W.~Hill, U.~Kuruganti, and K.~Englehart, ``Motion normalized proportional control for improved pattern recognition-based myoelectric control,'' \emph{IEEE Transactions on Neural Systems and Rehabilitation Engineering}, vol.~22, no.~1, pp. 149--157, 2014.

\bibitem{ganin2016dann}
Y.~Ganin, E.~Ustinova, H.~Ajakan, P.~Germain, H.~Larochelle, F.~Laviolette, M.~March, and V.~Lempitsky, ``Domain-{Adversarial} {Training} of {Neural} {Networks},'' \emph{Journal of Machine Learning Research}, vol.~17, no.~59, pp. 1--35, 2016.

\bibitem{levy2020large}
D.~Levy, Y.~Carmon, J.~C. Duchi, and A.~Sidford, ``Large-scale methods for distributionally robust optimization,'' 2020.

\bibitem{kumar2022finetuning}
A.~Kumar, A.~Raghunathan, R.~Jones, T.~Ma, and P.~Liang, ``Fine-tuning can distort pretrained features and underperform out-of-distribution,'' in \emph{Proc. Int. Conf. Learn. Represent. (ICLR)}, 2022.

\bibitem{campbell2020current}
E.~Campbell, A.~Phinyomark, and E.~Scheme, ``\BIBforeignlanguage{en}{Current {Trends} and {Confounding} {Factors} in {Myoelectric} {Control}: {Limb} {Position} and {Contraction} {Intensity}},'' \emph{\BIBforeignlanguage{en}{Sensors}}, vol.~20, no.~6, p. 1613, Mar. 2020.

\bibitem{Hargrove2007Realtime}
L.~Hargrove, Y.~Losier, B.~Lock, K.~Englehart, and B.~Hudgins, ``A real-time pattern recognition based myoelectric control usability study implemented in a virtual environment,'' in \emph{Proceedings of the 29th Annual International Conference of the IEEE Engineering in Medicine and Biology Society (EMBC)}, 2007, pp. 4842--4845.

\bibitem{campbell2024CIIL}
E.~Campbell, E.~Eddy, S.~Bateman \emph{et~al.}, ``Context-informed incremental learning improves both the performance and resilience of myoelectric control,'' \emph{Journal of NeuroEngineering and Rehabilitation}, vol.~21, p.~70, 2024.

\bibitem{campbell2020comparison}
E.~Campbell, J.~Chang, A.~Phinyomark, and E.~Scheme, ``A {Comparison} of {Amputee} and {Able}-{Bodied} {Inter}-{Subject} {Variability} in {Myoelectric} {Control},'' Mar. 2020, arXiv:2003.03481 [eess.SP].

\bibitem{hajian2024generalizing}
G.~Hajian, E.~Campbell, M.~Ansari, E.~Morin, A.~Etemad, K.~Englehart, and E.~Scheme, ``Generalizing upper limb force modeling with transfer learning: A multimodal approach using {EMG} and {IMU} for new users and conditions,'' \emph{IEEE Transactions on Neural Systems and Rehabilitation Engineering}, vol.~32, pp. 391--400, 2024.

\end{thebibliography}

\end{document}